\documentclass[prl,aps,twocolumn]{revtex4-2}
\usepackage{graphicx}
\usepackage{natbib}
\usepackage{xcolor}

\usepackage{subfig}
\usepackage{float}
\usepackage{amsmath}
\usepackage{hyperref}

\begin{document}

\title{Neutron-nucleus  scattering in an  $L^2$-integrable basis}

\author{Bui Minh Loc}
\affiliation{ San Diego State University,
5500 Campanile Drive, San Diego, CA 92182} 

\author{Calvin W. Johnson}
\affiliation{ San Diego State University,
5500 Campanile Drive, San Diego, CA 92182} 

\begin{abstract}
First-principles calculations of scattering off quantum many-body systems such as atoms and nuclei is important but nontrivial, requiring large-dimension ($10^3$ to $10^8$ or more) bases. For the first time we apply to many-body targets a recently developed approach to find phase shifts in $L^2$-integrable bases, computing neutron-nucleus scattering using the same interactions as \textit{ab initio} models of nuclei, in very small (dimension 3-6) bases. Our work shows the mean field dominates low-energy nuclear scattering; it also provides a foundation for more efficient future methodologies.
\end{abstract}

\maketitle

One of the main paths to investigating the quantum world has been through scattering and reactions. 
Yet coupling scattering states to detailed microscopic models of atoms and nuclei is a challenge~\cite{ptasinska2022electron,johnson2020white}.

One can replace the complicated many-body 
target by a one-body potential, for example by by considering the self-consistent 
mean-field (SCMF) potential, where the Pauli principle is respected through explicit antisymmetriziation; in nuclear physics, this has been done using phenomenological density functionals~\cite{VAUTHERIN1968552,DOVER1971559,DOVER1972373,AnhPRC1062022,AnhPRC106L2022}.  One can also  include many-body correlations 
via the Feshbach formalism~\cite{FESHBACH1958357,FESHBACH1962287} to obtain an effective potential, 
also called an optical potential.  Although there has been  
efforts to derive optical potentials from both SCMF~\cite{OsmanJPG1981,QingbiaoZPA1981,ShenPRC2009,PilipenkoPRC2010,PilipenkoPRC2012} and from more fundamental pictures~\cite{PhysRevC.95.024315,hebborn2023optical,wtmw-b26w},
the most widely used such potentials are  phenomenological~\cite{koning2003local,PhysRevC.107.014602}. 


To explicitly represent many-body systems, one can turn to  configuration-interaction 
methods, expanding eigenstates in a many-body bases~\cite{shavitt1998history,CaurierSMreview2005}. In the  nuclear no-core shell model (NCSM)~\cite{barrett2013ab}
and its extensions, one  uses interaction matrix elements fitted carefully 
to few-body data including nucleon-nucleon scattering, providing a realistic and rigorous foundation. As the many-body basis is built from 
$L^2$-integrable functions, such as isotropic harmonic oscillator (HO) states, 
coupling to the continuum is nontrivial, inspiring extensions 
that explicitly account for the continuum ~\cite{BENNACEUR1999289,BENNACEUR200075,
VolyaVZPRC2006,MichelPRL892002}. 



The NCSM extended via 
  the resonating group method,  (NCSM-RGM)~\cite{PhysRevLett.101.092501,PhysRevC.79.044606,PhysRevC.82.034609} as well as  a 
fuller description of the continuum (NCSMC)~\cite{PhysRevC.87.034326} computes 
the potential between the target and the projectile, often 
including different breakup channels, and then solves the 
resulting scattering equations~\cite{tang1978resonating}.
 Alternately, and relevant to this Letter, 
the $J$-matrix approach~\cite{PhysRevA.9.1201,alhaidari2008j} builds upon the insight 
that in a basis of HO states (or, in principle, 
more general states~\cite{PhysRevA.64.042703}), the representation of the kinetic energy 
is a tridiagonal or Jacobi matrix. In this representation, 
the standard $N_\mathrm{max}$ truncation for the NCSM 
corresponds simply to a boundary condition in HO space. This leads to the 
single-state harmonic oscillator representation of the 
Schr\"odinger equation, or SS-HORSE~\cite{PhysRevC.94.064320}. 
The SS-HORSE, NCSM-RGM and NCSMC extract phase shifts and other scattering parameters by matching the wave function asymptotically, that is, 
sufficiently ``far'' that the interaction with the target vanishes and 
the wave function is a phase-shifted free scattering state. 
This requirement leads to very large NCSM calculations, with dimensions up to $10^8$ or more.


\begin{figure}[htb]
    \centering
    \includegraphics[width=1.0\linewidth]{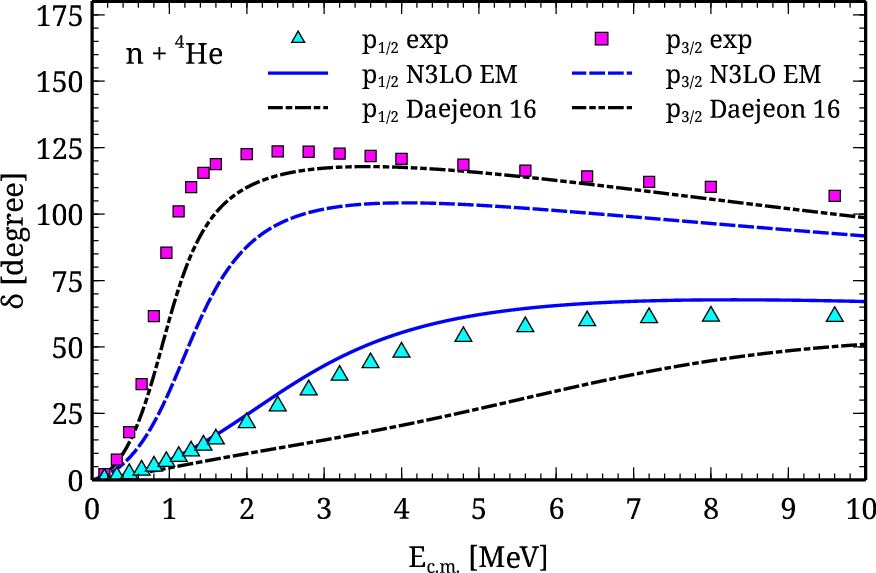}
    \caption{$n+^{4}$He $p$-wave phase shifts with realistic \textit{NN} interactions:    N$^3$LO Entem-Machleidt (dimension = 6 at $\hbar \Omega=17$ MeV) and Daejeon16  (dimension = 4 at $\hbar \Omega=17.5$ MeV, due to file availability). }
    \label{fig:nhe4}
\end{figure}

An alternative to matching, however, is to use scattering overlap relations~\cite{pinkston1965form,timofeyuk1998one,rodberg1967introduction,berggren1965overlap,PhysRevC.86.044330,nd89-xtlw}.
In coordinate space, overlap relations are integrals, akin to  
theorems in calculus relating integrals over the volume to values 
on the boundary.  Such overlap relations,  dominated by 
the ``volume,''  are much less sensitive than matching to  errors in the 
asymptotic region or in the scattering state itself~\cite{flores2022variational}. Recently an extension of overlap 
relations to the $J$-matrix method, e.g., in a HO basis, 
has been introduced, with applications to proton-neutron scattering~\cite{nd89-xtlw}. 
In this Letter, we apply these $J$-matrix scattering overlap relations to 
elastic scattering neutrons off $^4$He, $^{12}$C, $^{16}$O, and $^{40}$Ca, 
using well-established \textit{ab initio} nucleon-nucleon interactions. By using closed-shell targets, we can use simplified models 
of the wave functions, so that the problem becomes single-channel 
potential scattering, including an exact accounting for the Pauli exclusion principle. Despite the simplicity of our approach,  agreement with experimental data is far better than one might  expect. 
Because we use  robust scattering overlap relations and employ 
the Lippmann-Schwinger equation to obtain scattering solutions at arbitrary 
energies, we are able to work in very small model spaces, much smaller than the NCSM calculations. Our results demonstrate 
 that, starting from a \textit{ab initio} interaction fitted to two- and few-body data, the resulting mean-field potential  captures much of the scattering physics. Furthermore, our approach 
 is a good candidate for future work with more detailed models.

\textbf{Methods.} Any calculation of scattering has two parts: 
generating the scattering state(s), and extracting from the scattering 
state the phase shifts or the scattering or $S$-matrix.  Here 
we consider single-channel scattering and  only extract 
phase shifts, but the $J$-matrix method can be extended to 
coupled channels~\cite{broad1976j,PhysRevA.76.062706}.

We work in a basis of harmonic oscillator (HO) states,  $\{ | \phi_n \rangle \}$, where $n$ is the radial nodal quantum number; other quantum numbers such as orbital angular momentum $l$, etc., are implied but not written down explicitly. (The $J$-matrix methods can be extended to other $L^2$-integrable bases~\cite{PhysRevA.64.042703}.)
Wave functions, such as positive-energy solutions to the 
Schr\"odinger equation $\hat{H}| u \rangle = E | u \rangle$, 
are expanded in this basis,
\begin{equation}
    | u \rangle = \sum_{n=0} u_n | \phi_n \rangle.
\end{equation}
The Hamiltonian $\hat{H}$ is composed of the kinetic energy  $\hat{T}$
and the potential energy $\hat{V}$. When $\hat{V}$ vanishes, we have free scattering solutions, such 
as the solution is regular at the origin (in coordinate space).
$ | f \rangle = \sum_n f_n | \phi_n \rangle $. The irregular solution is more subtle, because 
HO basis states are  regular at the origin. Hence, one uses a modified `irregular' 
solution $| \tilde{g} \rangle = \sum_n \tilde{g}_n | \phi_n \rangle$,  regular at the origin but asymptotically recovering the 
irregular solution. One can rigorously define $ |\tilde{g} \rangle$ as the solution to an 
inhomogeneous differential equation~\cite{yamani1975j,nd89-xtlw}.  As the kinetic energy in the HO basis is 
tridiagonal and analytically known, the coefficients $f_n, \tilde{g}_n$ are known analytically. 

For single-channel scattering, one assumes  asymptotically (for large $n$), 
\begin{equation}
    u_n \rightarrow  f_n \cos \delta + \tilde{g}_n \sin \delta . \label{eq:asymp}
\end{equation}
This requires that the interaction between the target and the projectile becomes small at large $n$.
The most common way to extract the phase shift is through matching, that is, use Eq.~(\ref{eq:asymp}) 
explicitly. In the case of SS-HORSE, by truncating the model space one has a boundary 
condition that for some $n_\mathrm{max}$, $u_{n_\mathrm{max}}=0$.  Such a boundary condition 
discretizes the continuum, but limits the accessible scattering energies. 

Rather than matching, we use a scattering overlap relation in the HO space to
extract the phase shift~\cite{nd89-xtlw}:
\begin{equation}
    \tan \delta = -\frac{\sum_{m,n} f_m V_{mn} u_n}{\alpha_0 u_0 + \sum_{m,n} \tilde{g}_m V_{mn} u_n}. \label{SOR}
\end{equation}
Here  $V_{mn} = \langle m | \hat{V} | n \rangle$ is the matrix element of the potential in 
HO space. 
One can show  
$\alpha_0= W(f,g)/f_0$~\cite{nd89-xtlw}, where $W(f,g)$ is the Wronskian of the  regular and irregularly free scattering functions $f$ and $g$, respectively, and
that $\alpha_0$ is directly related to 
constructing $\tilde{g}_n$ via an inhomogeneous differential equation.

In order to use Eq.~(\ref{SOR}), we need the full scattering solution coefficients $u_n$ at different 
energies. Rather than use a boundary condition to discretize the continuum, we use the 
Lippmann-Schwinger equation, 
$    | u \rangle = | f \rangle + (E-\hat{H})^{-1} \hat{V} | f \rangle.
$
This can generate scattering solutions at arbitrary energies and  works remarkably well. 

\begin{figure}[htb]
    \centering
    \includegraphics[width=1.0\linewidth]{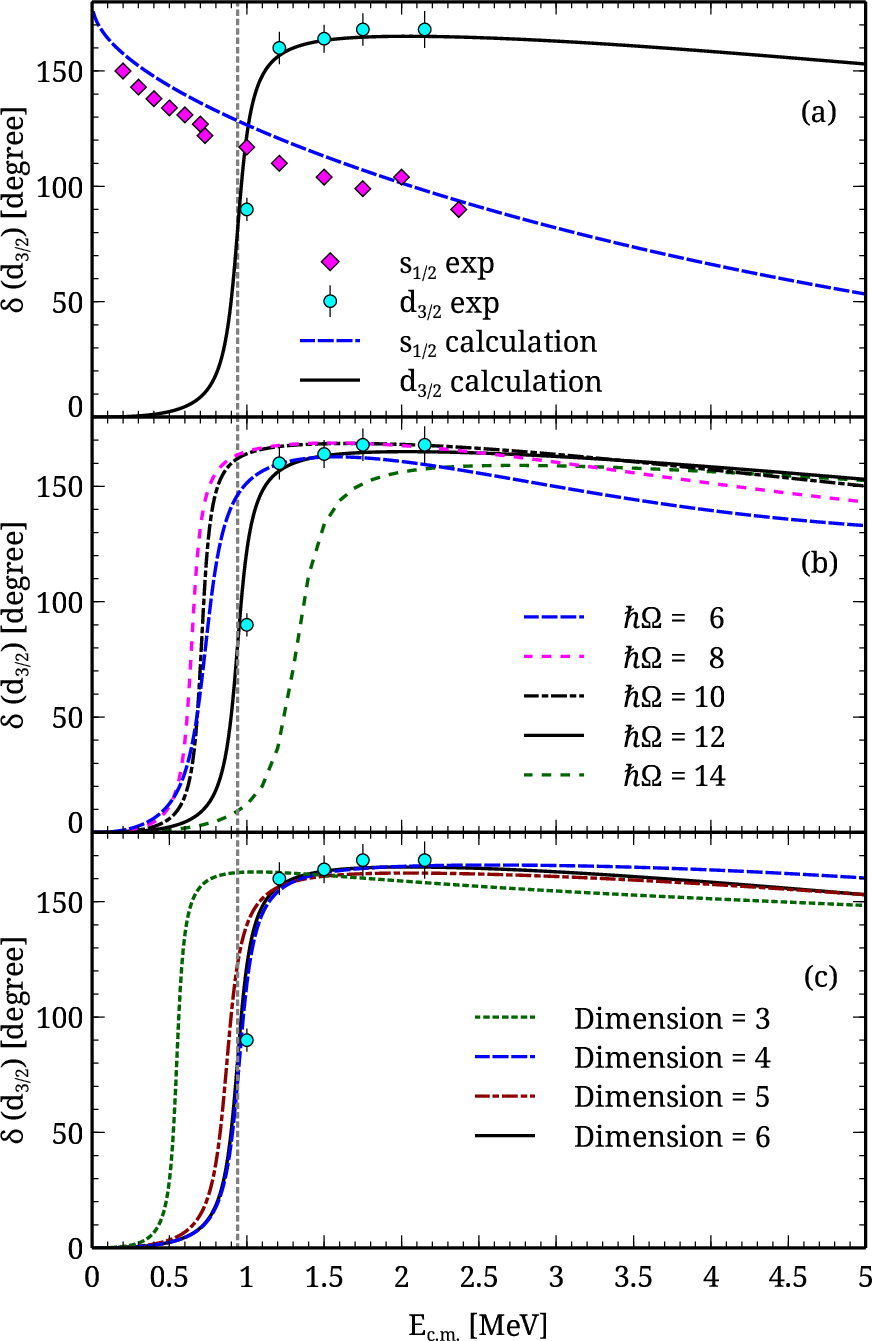}
    \caption{$n+^{16}\mathrm{O}$  phase shifts, with  Entem-Machleidt N3LO.
    Panel (a) shows the $s_{1/2}$ and $d_{3/2}$ phase shifts computed at $\hbar \Omega = 12$ MeV with  dimension = 6. Panel (b) gives the (dimension = 6) $d_{3/2}$ phase shifts for different values of $\hbar \Omega$, while Panel (c) gives the same for different values of $N_\mathrm{pot}$ (at $\hbar \Omega=12$ MeV).}
    \label{fig:d32no16}
\end{figure}

We now need the  the interaction. 
Let ${\cal V}_{mj,ni}$ be the antisymmetrized 
lab-frame matrix elements in a HO basis. 
 We obtain an effective potential by \textit{folding}~\cite{PhysRevC.66.014610,PhysRevC.94.034612} the two-body interaction with the one-body density matrix of the target $\rho_{ij}$,
\begin{equation}
    V_{mn} = \sum_{ij} {\cal V}_{mj,ni} \rho_{ij}. \label{Veffgeneral}
\end{equation}
By modeling the targets as filled HO shells, we have  $\rho_{ij} = \delta_{ij}$, where $i$ is an occupied single-particle state. Then,
\begin{equation}
    V_{mn} = \sum_{i \in \text{core}}{\cal V}_{mi,ni} \label{Vefffrozen}.
\end{equation}
Aside from using a `reduced' mass of the scattering system in the kinetic term, $M = A/(A-1)m_{\text{neutron}}$~\cite{ColoCPC2013},
we leave center-of-mass corrections to future work.  

The only parameters in these calculations are 
the basis HO length parameter $b= \sqrt{\hbar/M \Omega}$, where $\Omega$ is the oscillator frequency, and the size of the model space for the scattering neutron. 
Note that, despite working in an $L^2$-integrable basis, 
the kinetic energy is handled exactly within the model space, and that the 
main approximation is the truncation of the interaction energy matrix elements. We account for the 
Pauli principle by  excluding occupied orbitals. For example, 
for a $^4$He target, a scattered $s$-wave would be blocked from the $0s_{1/2}$ orbital.

For most of our calculations, we chose $\hbar\Omega$ so that the simple filled shells 
for the target  approximate the experimental charge radius, that is, values 
of 17, 14, 12, and 10 MeV for $^4$He, $^{12}$C, $^{16}$O, and $^{40}$Ca, respectively.  
(This is different from the standard NCSM practice of choosing $\hbar \Omega$ so as to 
minimize the energy~\cite{barrett2013ab}.)
Below (Fig.~\ref{fig:d32no16}(b)) we find  only a mild sensitivity to $\hbar \Omega$, although 
we confirm that the strategy of matching the radius is the correct one.
Note the  density profiles have Gaussian rather than exponential fall-offs.

\textbf{Results and discussion.} 
We use  two-body (only) interactions  from chiral effective field theory: 
the Entem-Machleidt interaction at next-to-next-to-next-to-leading order (N3LO)~\cite{PhysRevC.68.041001},  generated 
using the \texttt{NuHamil} code~\cite{Miyagi2023},  softened via the similarity renomalization 
group~\cite{hergert2016medium}, to a resolution parameter of $\lambda = 2\, \mathrm{fm}^{-1}$~\cite{MACHLEIDT20111, EPELBAUM2005362}; and the very soft interaction 
Daejeon16~\cite{shirokov2016n3lo}. 
Due to the availability of files for the latter, 
the corresponding model space  was smaller.

First, we consider neutron scattering off $^4$He.
This has been previously computed in NCSM-RGM using the Entem-Machleidt N3LO interaction~\cite{PhysRevLett.101.092501,PhysRevC.79.044606}, with the NCSM calculations having 
dimensions  from 96,340 to 10 million. It has also been  computed in the SS-HORSE $J$-matrix method, albeit via matching, using the 
Daejeon16 interaction~\cite{PhysRevC.94.064320}, with NCSM dimensions ranging from 3,944 to 643 million; because SS-HORSE uses a boundary condition to discretize the continuum,  large dimensions were necessary to reach lower energies. 
Coordinate-space 
integral  relations were employed in conjunction with Green's Function Monte Carlo calculations~\cite{q4dy-vhv1} using the Argonne V18 interaction~\cite{PhysRevC.51.38} to compute $n+\alpha$ phase shifts. None of these calculations used three-body forces, which affects 
spin-orbit splitting between $0p_{1/2}$ and $0p_{3/2}$ and thus could influence the phase shifts~\cite{PhysRevLett.101.092501,PhysRevC.79.044606}. 

 Fig.~\ref{fig:nhe4} shows our $p$-wave $n+^4$He phase shifts,  similar in quality to those 
 of the NCSM-RGM and SS-HORSE calculations, but in  model space dimensions of only 
 6 and 4 for Entem-Machleidt and Daejeon16, respectively.
The small dimensions are possible because of the insensitivity of scattering overlap relations to tails of the wave functions.  The 
Lippmann-Schwinger equation provides solutions at arbitrary  scattering energies.

\begin{figure}[htb]
\includegraphics[width=1.0\linewidth]{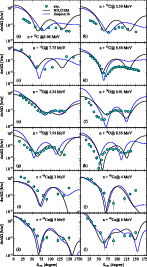}
    \caption{Differential cross-sections for neutron scattering from $^{12}$C, $^{16}$O, and $^{40}$Ca up to 10 MeV. The calculations within the frozen-core approximation using N$^3$LO Entem-Machleidt are the (black) solid lines, and using Daejeon 16 are  (blue) dashed lines.  Data taken from \cite{Glasgow1976_n12Cusedexp} for $^{12}$C target, \cite{Kinney1972_n16Ousedexp} for $^{16}$O , and \cite{Topke1974_n40Causedexp} for $^{40}$Ca.}
    \label{fig:DAallinone}
\end{figure}

We explore  our methodology via $n+^{16}$O scattering, using 
the Entem-Machleidt N3LO interaction,
and compare our $s_{1/2}$ and $d_{3/2}$
channel results against experiment~\cite{DOVER1971559}.
Fig.~\ref{fig:d32no16}(a) shows  scattering phase shifts calculated with dimensions of 6 at $\hbar \Omega = 12$ MeV, which approximately reproduces the nuclear radius;
we obtain the energy resonance $E_R = 0.95$ MeV and $\Gamma = 141.1$ keV.
 At  $\hbar \Omega = 8$ and $10$, shown in Fig.~\ref{fig:d32no16} (b), we obtain $E_R = 0.65, 0.71$ MeV and $\Gamma = 96.7, 89.7$ keV, respectively. 
For comparison, the  experimental  $d_{3/2}$ resonance is located at $E_R = 0.94$ MeV above the neutron emission threshold, corresponding to the $3/2^+$ state at 5.09 MeV in  $^{17}$O, with a  width of $\Gamma = 96(5)$ keV (Ref.~\cite{Tilley1993A16A17_nO16d32usedexp}, Table 17.10, page 24).
Fig.~\ref{fig:d32no16}  (c) compares
experimental $d_{3/2}$ phase shifts   at different scattering state dimensions. Even with 
a dimension of only 3, we obtain the 
correct qualitative behavior, and there is  little difference among
 dimensions $\geq 4$.


Lastly, we show  differential angular cross-sections in Fig.~\ref{fig:DAallinone}. Although the present calculations employ the simplest frozen-core approximation, they already reproduce many qualitative features of neutron elastic scattering from $^{12}$C, $^{16}$O, and $^{40}$Ca.  Scattering at large angles requires the proper treatment of open channels, which in the optical model corresponds to the imaginary potential, so
given the simplicity of our target it is not surprising that the large scattering angle cross-sections have poorer agreement with experiment~\cite{PhysRevC.33.1826}. Furthermore, our nonphysical, Gaussian falloff may partially account for the misplacement of the diffraction minima. Nonetheless, we capture many of the experimental features.

Not shown are the calculated $s_{1/2}$ phase shifts for $n+^4$He and $p$-wave
phase shifts for $n+^{16}$O, which 
agree poorly with experiment. Because 
these are just below the Fermi surface,  these phase 
shifts are  likely sensitive to 
elements ignored in our severe 
approximation, such as approximating
filled orbitals as HO states.

\textbf{Conclusions.}
The  significance of our results lies not in the frozen-core approximation itself, but in demonstrating that,  
starting from realistic interactions, low-energy scattering in a mean-field potential can largely reproduces experiment. We identify  features requiring future improvement, such as scattering at large angles.

Furthermore, we have shown that
neutron-nucleus scattering can be formulated entirely within the same HO basis framework used in modern shell-model calculations at low computational cost. 
The present formalism thus offers a straightforward path toward more realistic descriptions. The frozen one-body density can be replaced by correlated densities obtained from the NCSM or other many-body calculations.
While we have applied 
this approach to nucleon-nucleus scattering, it could be applied to other  systems such as electron-atom scattering, etc..

Near future steps will 
address charged-particle scattering, as well as generalizing to more complex target wave functions. In the latter case the formalism will likely be  similar to that of the NCSM-RGM and the NCSMC, albeit using the Lippmann-Schwinger equation to generate scattering states and employing overlap relations rather than matching.
Long-term work would naturally address composite projectiles such as deuterons and alpha particles, as well as transfer reactions, etc..

\textit{Acknowledgements.}  Conversations with Kenneth Nollett have been  fruitful in the development of this work. CWJ acknowledges  helpful discussion of 
preliminary results at the May, 2026 workshop on ``Nuclear physics for astrophysics,'' at the ECT* in Trento, Italy. We thank J. Vary and M. Caprio for access to Daejeon16 files.
This material is based upon work supported by the U.S. Department of Energy, under Award Number DE-NA0004075.

\bibliographystyle{unsrt}
\bibliography{jmatrix}

\begin{thebibliography}{10}

\bibitem{ptasinska2022electron}
Sylwia Ptasinska, Marcio T do~N Varella, Murtadha~A Khakoo, Daniel~S Slaughter,
  and Stephan Denifl.
\newblock Electron scattering processes: fundamentals, challenges, advances,
  and opportunities.
\newblock {\em The European Physical Journal D}, 76(10):179, 2022.

\bibitem{johnson2020white}
Calvin~W Johnson, Kristina~D Launey, Naftali Auerbach, Sonia Bacca, Bruce~R
  Barrett, Carl~R Brune, Mark~A Caprio, Pierre Descouvemont, WH~Dickhoff,
  Charlotte Elster, et~al.
\newblock White paper: from bound states to the continuum.
\newblock {\em Journal of Physics G: Nuclear and Particle Physics},
  47(12):123001, 2020.

\bibitem{VAUTHERIN1968552}
D.~Vautherin and M.~Vénéroni.
\newblock Potential scattering by a nuclear {H}artree-{F}ock field.
\newblock {\em Physics Letters B}, 26(9):552--555, 1968.

\bibitem{DOVER1971559}
Carl~B. Dover and Nguyen {Van Giai}.
\newblock Low-energy neutron scattering by a {H}artree-{F}ock field.
\newblock {\em Nuclear Physics A}, 177(2):559--576, 1971.

\bibitem{DOVER1972373}
C.B. Dover and Nguyen {Van Giai}.
\newblock The nucleon-nucleus potential in the {H}artree-{F}ock approximation
  with {S}kyrme's interaction.
\newblock {\em Nuclear Physics A}, 190(2):373--400, 1972.

\bibitem{AnhPRC1062022}
Nguyen Le~Anh and Bui Minh~Loc.
\newblock Low-energy $^{7}\mathrm{Li}(n,\ensuremath{\gamma})^{8}\mathrm{Li}$
  and $^{7}\mathrm{Be}(p,\ensuremath{\gamma})^{8}\mathrm{B}$ radiative capture
  reactions within the {S}kyrme {H}artree-{F}ock approach.
\newblock {\em Phys. Rev. C}, 106:014605, Jul 2022.

\bibitem{AnhPRC106L2022}
Nguyen Le~Anh, Bui~Minh Loc, Naftali Auerbach, and Vladimir Zelevinsky.
\newblock Single-particle properties of the near-threshold proton-emitting
  resonance in $^{11}\mathrm{B}$.
\newblock {\em Phys. Rev. C}, 106:L051302, Nov 2022.

\bibitem{FESHBACH1958357}
Herman Feshbach.
\newblock Unified theory of nuclear reactions.
\newblock {\em Annals of Physics}, 5(4):357--390, 1958.

\bibitem{FESHBACH1962287}
Herman Feshbach.
\newblock A unified theory of nuclear reactions. {II}.
\newblock {\em Annals of Physics}, 19(2):287--313, 1962.

\bibitem{OsmanJPG1981}
A~Osman, M~Y Ismail, and M~M Osman.
\newblock Optical-model potential using a generalised {S}kyrme force.
\newblock {\em Journal of Physics G: Nuclear Physics}, 7(3):347, mar 1981.

\bibitem{QingbiaoZPA1981}
Shen Qingbiao, Zhang Jingshang, Tian Ye, Ma~Zhongyu, and Zhuo Yizhong.
\newblock Semi-microscopic optical potential calculation by the nuclear matter
  approach.
\newblock {\em Zeitschrift f{\"u}r Physik A Atoms and Nuclei}, 303(1):69--83,
  Mar 1981.

\bibitem{ShenPRC2009}
Qing-biao Shen, Yin-lu Han, and Hai-rui Guo.
\newblock Isospin dependent nucleon-nucleus optical potential with {S}kyrme
  interactions.
\newblock {\em Phys. Rev. C}, 80:024604, Aug 2009.

\bibitem{PilipenkoPRC2010}
V.~V. Pilipenko, V.~I. Kuprikov, and A.~P. Soznik.
\newblock Skyrme interaction and elastic nucleon-nucleus scattering in the
  optical model.
\newblock {\em Phys. Rev. C}, 81:044614, Apr 2010.

\bibitem{PilipenkoPRC2012}
V.~V. Pilipenko and V.~I. Kuprikov.
\newblock Extended skyrme interaction in the microscopic optical model of
  nucleon-nucleus scattering.
\newblock {\em Phys. Rev. C}, 86:064613, Dec 2012.

\bibitem{PhysRevC.95.024315}
J.~Rotureau, P.~Danielewicz, G.~Hagen, F.~M. Nunes, and T.~Papenbrock.
\newblock Optical potential from first principles.
\newblock {\em Phys. Rev. C}, 95:024315, Feb 2017.

\bibitem{hebborn2023optical}
C~Hebborn, FM~Nunes, G~Potel, WH~Dickhoff, JW~Holt, MC~Atkinson, RB~Baker,
  C~Barbieri, G~Blanchon, M~Burrows, et~al.
\newblock Optical potentials for the rare-isotope beam era.
\newblock {\em Journal of Physics G: Nuclear and Particle Physics},
  50(6):060501, 2023.

\bibitem{wtmw-b26w}
G.~H. Sargsyan, G.~Potel, K.~Kravvaris, and J.~E. Escher.
\newblock Microscopic optical potentials from a green's function approach.
\newblock {\em Phys. Rev. C}, 112:054606, Nov 2025.

\bibitem{koning2003local}
AJ~Koning and JP~Delaroche.
\newblock Local and global nucleon optical models from 1 kev to 200 {M}ev.
\newblock {\em Nuclear physics A}, 713(3-4):231--310, 2003.

\bibitem{PhysRevC.107.014602}
C.~D. Pruitt, J.~E. Escher, and R.~Rahman.
\newblock Uncertainty-quantified phenomenological optical potentials for
  single-nucleon scattering.
\newblock {\em Phys. Rev. C}, 107:014602, Jan 2023.

\bibitem{shavitt1998history}
Isaiah Shavitt.
\newblock The history and evolution of configuration interaction.
\newblock {\em Molecular Physics}, 94(1):3--17, 1998.

\bibitem{CaurierSMreview2005}
E.~Caurier, G.~Mart\'{\i}nez-Pinedo, F.~Nowacki, A.~Poves, and A.~P. Zuker.
\newblock The shell model as a unified view of nuclear structure.
\newblock {\em Rev. Mod. Phys.}, 77:427--488, Jun 2005.

\bibitem{barrett2013ab}
Bruce~R Barrett, Petr Navr{\'a}til, and James~P Vary.
\newblock Ab initio no core shell model.
\newblock {\em Progress in Particle and Nuclear Physics}, 69:131--181, 2013.

\bibitem{BENNACEUR1999289}
K.~Bennaceur, F.~Nowacki, J.~Okołowicz, and M.~Płoszajczak.
\newblock Study of the $^{7}\mathrm{Be}(p,\gamma)^8\mathrm{B}$ and
  $^7\mathrm{Li}(n,\gamma)^8\mathrm{Li}$ capture reactions using the shell
  model embedded in the continuum.
\newblock {\em Nuclear Physics A}, 651(3):289--319, 1999.

\bibitem{BENNACEUR200075}
K~Bennaceur, N~Michel, F~Nowacki, J~Okołowicz, and M~Płoszajczak.
\newblock Shell model description of $^{16}\mathrm{O}(p,\gamma)^{17}\mathrm{F}$
  and $^{16}\mathrm{O}(p,p)^{16}\mathrm{O}$ reactions.
\newblock {\em Physics Letters B}, 488(1):75--82, 2000.

\bibitem{VolyaVZPRC2006}
Alexander Volya and Vladimir Zelevinsky.
\newblock Continuum shell model.
\newblock {\em Phys. Rev. C}, 74:064314, Dec 2006.

\bibitem{MichelPRL892002}
N.~Michel, W.~Nazarewicz, M.~P\l{}oszajczak, and K.~Bennaceur.
\newblock Gamow shell model description of neutron-rich nuclei.
\newblock {\em Phys. Rev. Lett.}, 89:042502, Jul 2002.

\bibitem{PhysRevLett.101.092501}
Sofia Quaglioni and Petr Navr\'atil.
\newblock Ab initio many-body calculations of
  $n\mathrm{\text{\ensuremath{-}}}^{3}\mathrm{H}$,
  $n\mathrm{\text{\ensuremath{-}}}^{4}\mathrm{He}$,
  $p\mathrm{\text{\ensuremath{-}}}^{3,4}\mathrm{He}$, and
  $n\mathrm{\text{\ensuremath{-}}}^{10}\mathrm{Be}$ scattering.
\newblock {\em Phys. Rev. Lett.}, 101:092501, Aug 2008.

\bibitem{PhysRevC.79.044606}
Sofia Quaglioni and Petr Navr\'atil.
\newblock Ab initio many-body calculations of nucleon-nucleus scattering.
\newblock {\em Phys. Rev. C}, 79:044606, Apr 2009.

\bibitem{PhysRevC.82.034609}
Petr Navr\'atil, Robert Roth, and Sofia Quaglioni.
\newblock Ab initio many-body calculations of nucleon scattering on
  $^{4}\mathrm{He}$, $^{7}\mathrm{Li}$, $^{7}\mathrm{Be}$, $^{12}\mathrm{C}$,
  and $^{16}\mathrm{O}$.
\newblock {\em Phys. Rev. C}, 82:034609, Sep 2010.

\bibitem{PhysRevC.87.034326}
Simone Baroni, Petr Navr\'atil, and Sofia Quaglioni.
\newblock Unified ab initio approach to bound and unbound states: No-core shell
  model with continuum and its application to ${}^{7}${H}e.
\newblock {\em Phys. Rev. C}, 87:034326, Mar 2013.

\bibitem{tang1978resonating}
Yau-Chien Tang, Mark LeMere, and DR~Thompsom.
\newblock Resonating-group method for nuclear many-body problems.
\newblock {\em Physics Reports}, 47(3):167--223, 1978.

\bibitem{PhysRevA.9.1201}
Eric~J. Heller and Hashim~A. Yamani.
\newblock New ${L}^{2}$ approach to quantum scattering: Theory.
\newblock {\em Phys. Rev. A}, 9:1201--1208, Mar 1974.

\bibitem{alhaidari2008j}
Abdulaziz~D Alhaidari, Eric~J Heller, Hashim~A Yamani, and Mohamed~S
  Abdelmonem.
\newblock The {J}-matrix method.
\newblock {\em Development and Applications (Springer, Berlin, 2008)}, 2008.

\bibitem{PhysRevA.64.042703}
H.~A. Yamani, A.~D. Alhaidari, and M.~S. Abdelmonem.
\newblock J-matrix method of scattering in any ${L}^{2}$ basis.
\newblock {\em Phys. Rev. A}, 64:042703, Sep 2001.

\bibitem{PhysRevC.94.064320}
A.~M. Shirokov, A.~I. Mazur, I.~A. Mazur, and J.~P. Vary.
\newblock Shell model states in the continuum.
\newblock {\em Phys. Rev. C}, 94:064320, Dec 2016.

\bibitem{pinkston1965form}
WT~Pinkston and GR~Satchler.
\newblock Form factors for nuclear stripping reactions.
\newblock {\em Nuclear Physics}, 72(3):641--656, 1965.

\bibitem{timofeyuk1998one}
NK~Timofeyuk.
\newblock One nucleon overlap integrals for light nuclei.
\newblock {\em Nuclear Physics A}, 632(1):19--38, 1998.

\bibitem{rodberg1967introduction}
Leonard~S Rodberg, Roy~M Thaler, and Raphael~Morton Thaler.
\newblock {\em Introduction to the quantum theory of scattering}, volume~26.
\newblock Academic Press, 1967.

\bibitem{berggren1965overlap}
Tore Berggren.
\newblock Overlap integrals and single-particle wave functions in direct
  interaction theories.
\newblock {\em Nuclear Physics}, 72(2):337--351, 1965.

\bibitem{PhysRevC.86.044330}
Kenneth~M. Nollett.
\newblock Ab initio calculations of nuclear widths via an integral relation.
\newblock {\em Phys. Rev. C}, 86:044330, Oct 2012.

\bibitem{nd89-xtlw}
Calvin~W. Johnson, Bui~Minh Loc, Austin Keller, and Kenneth~M. Nollett.
\newblock Scattering phase shifts from overlap relations in the ${J}$-matrix
  method.
\newblock {\em Phys. Rev. C}, 113:024004, Feb 2026.

\bibitem{flores2022variational}
Abraham~R. Flores and Kenneth~M. Nollett.
\newblock Variational {M}onte {C}arlo calculations of $n+^{3}\mathrm{H}$
  scattering.
\newblock {\em Phys. Rev. C}, 108:034001, Sep 2023.

\bibitem{broad1976j}
John~T Broad and William~P Reinhardt.
\newblock J-matrix method: multichannel scattering and photoionization.
\newblock {\em Journal of Physics B: Atomic and Molecular Physics}, 9(9):1491,
  1976.

\bibitem{PhysRevA.76.062706}
S.~A. Zaytsev.
\newblock ${J}$-matrix inverse-scattering approach for coupled channels with
  different thresholds.
\newblock {\em Phys. Rev. A}, 76:062706, Dec 2007.

\bibitem{yamani1975j}
Hashim~A Yamani and Louis Fishman.
\newblock {J}- matrix method: Extensions to arbitrary angular momentum and to
  {C}oulomb scattering.
\newblock {\em Journal of Mathematical Physics}, 16(2):410--420, 1975.

\bibitem{PhysRevC.66.014610}
L.~C. Chamon, B.~V. Carlson, L.~R. Gasques, D.~Pereira, C.~De~Conti, M.~A.~G.
  Alvarez, M.~S. Hussein, M.~A. C\^andido~Ribeiro, E.~S. Rossi, and C.~P.
  Silva.
\newblock Toward a global description of the nucleus-nucleus interaction.
\newblock {\em Phys. Rev. C}, 66:014610, Jul 2002.

\bibitem{PhysRevC.94.034612}
Dao~T. Khoa, Nguyen~Hoang Phuc, Doan~Thi Loan, and Bui~Minh Loc.
\newblock Nuclear mean field and double-folding model of the nucleus-nucleus
  optical potential.
\newblock {\em Phys. Rev. C}, 94:034612, Sep 2016.

\bibitem{ColoCPC2013}
Gianluca Colò, Ligang Cao, Nguyen {Van Giai}, and Luigi Capelli.
\newblock Self-consistent {RPA} calculations with {S}kyrme-type interactions:
  The skyrme\_rpa program.
\newblock {\em Computer Physics Communications}, 184(1):142--161, 2013.

\bibitem{PhysRevC.68.041001}
D.~R. Entem and R.~Machleidt.
\newblock Accurate charge-dependent nucleon-nucleon potential at fourth order
  of chiral perturbation theory.
\newblock {\em Phys. Rev. C}, 68:041001(R), Oct 2003.

\bibitem{Miyagi2023}
Takayuki Miyagi.
\newblock Nuhamil : A numerical code to generate nuclear two- and three-body
  matrix elements from chiral effective field theory.
\newblock {\em The European Physical Journal A}, 59(7):150, Jul 2023.

\bibitem{hergert2016medium}
Heiko Hergert, Scott~K Bogner, Titus~D Morris, Achim Schwenk, and Koshiroh
  Tsukiyama.
\newblock The in-medium similarity renormalization group: A novel ab initio
  method for nuclei.
\newblock {\em Physics reports}, 621:165--222, 2016.

\bibitem{MACHLEIDT20111}
R.~Machleidt and D.R. Entem.
\newblock Chiral effective field theory and nuclear forces.
\newblock {\em Physics Reports}, 503(1):1--75, 2011.

\bibitem{EPELBAUM2005362}
E.~Epelbaum, W.~Glöckle, and Ulf-G. Meißner.
\newblock The two-nucleon system at next-to-next-to-next-to-leading order.
\newblock {\em Nuclear Physics A}, 747(2):362--424, 2005.

\bibitem{shirokov2016n3lo}
AM~Shirokov, IJ~Shin, Y~Kim, M~Sosonkina, P~Maris, and JP~Vary.
\newblock {N3LO NN} interaction adjusted to light nuclei in \textit{ab exitu}
  approach.
\newblock {\em Physics Letters B}, 761:87--91, 2016.

\bibitem{q4dy-vhv1}
Abraham~R. Flores, Kenneth~M. Nollett, and Maria Piarulli.
\newblock Quantum monte carlo calculations of neutron-$\ensuremath{\alpha}$
  scattering via an integral relation.
\newblock {\em Phys. Rev. C}, 112:014008, Jul 2025.

\bibitem{PhysRevC.51.38}
R.~B. Wiringa, V.~G.~J. Stoks, and R.~Schiavilla.
\newblock Accurate nucleon-nucleon potential with charge-independence breaking.
\newblock {\em Phys. Rev. C}, 51:38--51, Jan 1995.

\bibitem{Glasgow1976_n12Cusedexp}
D.~W. Glasgow, F.~O. Purser, H.~Hogue, J.~C. Clement, K.~Stelzer, G.~Mack,
  J.~R. Boyce, D.~H. Epperson, S.~G. Buccino, P.~W. Lisowski, S.~G.
  Glendinning, E.~G. Bilpuch, H.~W. Newson, and C.~R. Gould.
\newblock Differential elastic and inelastic scattering of 9- to 15-{M}ev
  neutrons from carbon.
\newblock {\em Nuclear Science and Engineering}, 61(4):521--533, 1976.

\bibitem{Kinney1972_n16Ousedexp}
W.~E. Kinney and F.~G. Perey.
\newblock Neutron elastic and inelastic scattering from {O}-16.
\newblock Technical Report ORNL-4780, Oak Ridge National Laboratory, Oak Ridge,
  Tennessee, 1972.

\bibitem{Topke1974_n40Causedexp}
Rolf T{\"o}pke.
\newblock Messung und {R}esonanzanalyse von differentiellen elastischen
  {S}treuquerschnitten an $^{40}${C}a.
\newblock Technical Report KFK-2122, Kernforschungszentrum Karlsruhe,
  Karlsruhe, Germany, 1974.

\bibitem{Tilley1993A16A17_nO16d32usedexp}
D.~R. Tilley, H.~R. Weller, and C.~M. Cheves.
\newblock Energy levels of light nuclei $a=16$--17.
\newblock {\em Nuclear Physics A}, 564:1--183, 1993.

\bibitem{PhysRevC.33.1826}
J.~P. Delaroche, M.~S. Islam, and R.~W. Finlay.
\newblock Giant resonance coupling and l-dependent potentials for
  $^{16}\mathrm{O}$.
\newblock {\em Phys. Rev. C}, 33:1826--1829, May 1986.

\end{thebibliography}

\end{document}